\documentclass[preprint,aps,showpacs,prx,superscriptaddress,floatfix,10pt,footinbib]{revtex4-2}
\usepackage[T1]{fontenc}
\usepackage[utf8]{inputenc}
\usepackage{amsmath}
\usepackage{amssymb}
\usepackage{esint}
\usepackage{graphicx}
\usepackage{tabularx}
\usepackage{multirow}
\usepackage{bbold}
\usepackage{appendix}
\usepackage{bibunits}
\usepackage{verbatim}
\usepackage{bm}
\usepackage{xr-hyper}
\usepackage{lipsum} 
\usepackage{array}
\usepackage{comment}
\usepackage[table]{xcolor}
\usepackage{hyperref}
\usepackage{orcidlink}

\renewcommand{\selectlanguage}[1]{}

\newcolumntype{L}[1]{>{\raggedright\arraybackslash}p{#1}}

\makeatletter

\usepackage[table]{xcolor}
\usepackage{color}

\usepackage{physics}

\def\@fnsymbol#1{\ensuremath{\ifcase#1\or \dagger\or \ddagger\or
   \mathsection\or \mathparagraph\or \|\or **\or \dagger\dagger
   \or \ddagger\ddagger \else\@ctrerr\fi}}
    \makeatother

\makeatother

\usepackage{enumitem}
\usepackage{booktabs}
\usepackage{array,makecell}
\usepackage{tcolorbox,lipsum}

\newtcolorbox{FullWidthTwoColBox}{
    width=\textwidth,
    colback=gray!5,
    colframe=black,
    boxrule=0.8pt,
    arc=0pt,
    left=6pt,
    right=6pt,
    top=6pt,
    bottom=6pt
}

\usepackage{xcolor}

\begin{document}

\title{Quditology: harnessing many levels for quantum information processing}

\author{Abhijit Chakraborty \orcidlink{0000-0001-7487-9925}}
\thanks{These authors contributed equally to this work.}
\affiliation{Department of Physics, Virginia Tech, Blacksburg, VA 24061, USA}
\affiliation{Virginia Tech Center for Quantum Information Science and Engineering, Blacksburg, VA 24061, USA}
\affiliation{Department of Physics and Astronomy, Middle Tennessee State University, Murfreesboro, TN 37132, USA}

\author{Zheng Shi\,\orcidlink{0000-0003-3567-4059}}
\thanks{These authors contributed equally to this work.}
\affiliation{Institute for Quantum Computing, University of Waterloo, Waterloo, ON N2L 3G1, Canada}
\affiliation{Department of Physics \& Astronomy, University of Waterloo, Waterloo, ON N2L 3G1, Canada}

\author{Noah Goss\,\orcidlink{/0000-0002-3377-9415}}
\affiliation{Department of Physics, University of California, Berkeley, Berkeley, CA 94720, USA}
\affiliation{Computational Research Division, Lawrence Berkeley National Laboratory, Berkeley, CA 94720, USA}

\author{Irfan Siddiqi\,\orcidlink{0000-0003-2200-1090}}
\affiliation{Department of Physics, University of California, Berkeley, Berkeley, CA 94720, USA}
\affiliation{Computational Research Division, Lawrence Berkeley National Laboratory, Berkeley, CA 94720, USA}  

\author{Barry C. Sanders\,\orcidlink{0000-0002-8326-8912}}\affiliation{
Institute for Quantum Science and Technology, University of Calgary, AB T3A~0E1, Canada}
\affiliation{CERN, European Organisation for Nuclear Research, CH-1211 Geneva 23, Switzerland}

\author{Christine A. Muschik\,\orcidlink{0000-0002-4599-5107}}
\email{cmuschik@uwaterloo.ca}
\affiliation{Institute for Quantum Computing, University of Waterloo, Waterloo, ON N2L 3G1, Canada}
\affiliation{Department of Physics \& Astronomy, University of Waterloo, Waterloo, ON N2L 3G1, Canada}
\affiliation{Perimeter Institute for Theoretical Physics, Waterloo, ON N2L 2Y5, Canada}

\date{\today}

\def\thefootnote{*}\footnotetext{These authors contributed equally to this work.}

\begin{abstract}
~\\
Quantum computing offers the prospect of surpassing classical computation on certain tasks by leveraging quantum coherence and entanglement.
However, a significant gap remains between current quantum hardware and the requirements to transcend what classical computers can achieve, with decoherence limiting the capacity of storing quantum information and the number of logical steps that can be executed reliably.
In the long term, fault tolerance bridges this gap, but simply changing the numerical base from binary to higher base~$d$ is a viable way to `squeeze more juice'---get more quantum information processing---from a quantum computer of given size.
Additionally, for quantum simulations, qudits offer a natural encoding scheme for high-dimensional physical degrees of freedom.
Here, we review recent experimental progress in high-fidelity preparation, control, and measurement of qudits across various platforms such as trapped ions, superconducting circuits, solid-state nuclear spins, photonic systems, and neutral atoms.
We highlight early demonstrations of qudit-based quantum simulation, computation and error suppression enabled by these developments, heralding the transition of qudits from conceptual constructs to viable building blocks for quantum technologies.
Finally, we outline key challenges and potential future directions for scalable qudit quantum information processing.
\end{abstract}

\maketitle

\section*{Key points}

\begin{itemize}
    \item \textbf{Beyond binary}: The emerging field of qudit technology challenges the binary-logic paradigm borrowed from the highly successful binary paradigm of classical computing.
    \item \textbf{Extracting more computational power from quantum systems}: Quantum information processing with qudits harnesses more of the quantum resources provided by nature, albeit at the cost of increased control complexity.
    \item \textbf{Advances in quantum control result in a new wave of qudit capabilities}: The ability to control multi-level quantum systems has reached a point at which computing using qudits becomes feasible.
    \item \textbf{Diverse platforms advance qudit technology}: Various complementary platforms demonstrate qudit technology, mitigating the risk of delays due to technological roadblocks, in the same way that multiple binary quantum computer platforms has mitigated slowdowns in quantum computing progress.
\end{itemize}

\noindent\makebox[\textwidth]{\dotfill}

\section{Introduction}
Quantum information science is leading to beneficial disruptive advances for computing,
which converts some intractable problems to being feasible on quantum computers at scale~\cite{steane_quantum_1998}. Quantum computing has emerged over the past three decades as one of the most strategically important areas of technology
with numerous companies selling or providing on-cloud quantum computer services~\cite{macquarrie_emerging_2020,gill_quantum_2022}. The importance of quantum computing is exemplified by landmark algorithms such as 
Shor's factorisation algorithm~\cite{shor_polynomialtime_1997},
which has a deleterious impact on public-key cryptography, 
and quantum search for the function-inversion problem with a provable quadratic advantage in query complexity~\cite{grover_fast_1996}.
Furthermore, quantum heuristics such as quantum annealing~\cite{yarkoni_quantum_2022} and variational quantum algorithms~\cite{cerezo_variational_2021} could be advantageous for solving classically hard optimisation problems and estimating ground states and their energies, with applications to chemistry~\cite{cao_quantum_2019} and many-body physics~\cite{fauseweh_quantum_2024}.

\begin{figure}[ht]
    \centering
    \includegraphics[width=0.53\linewidth]{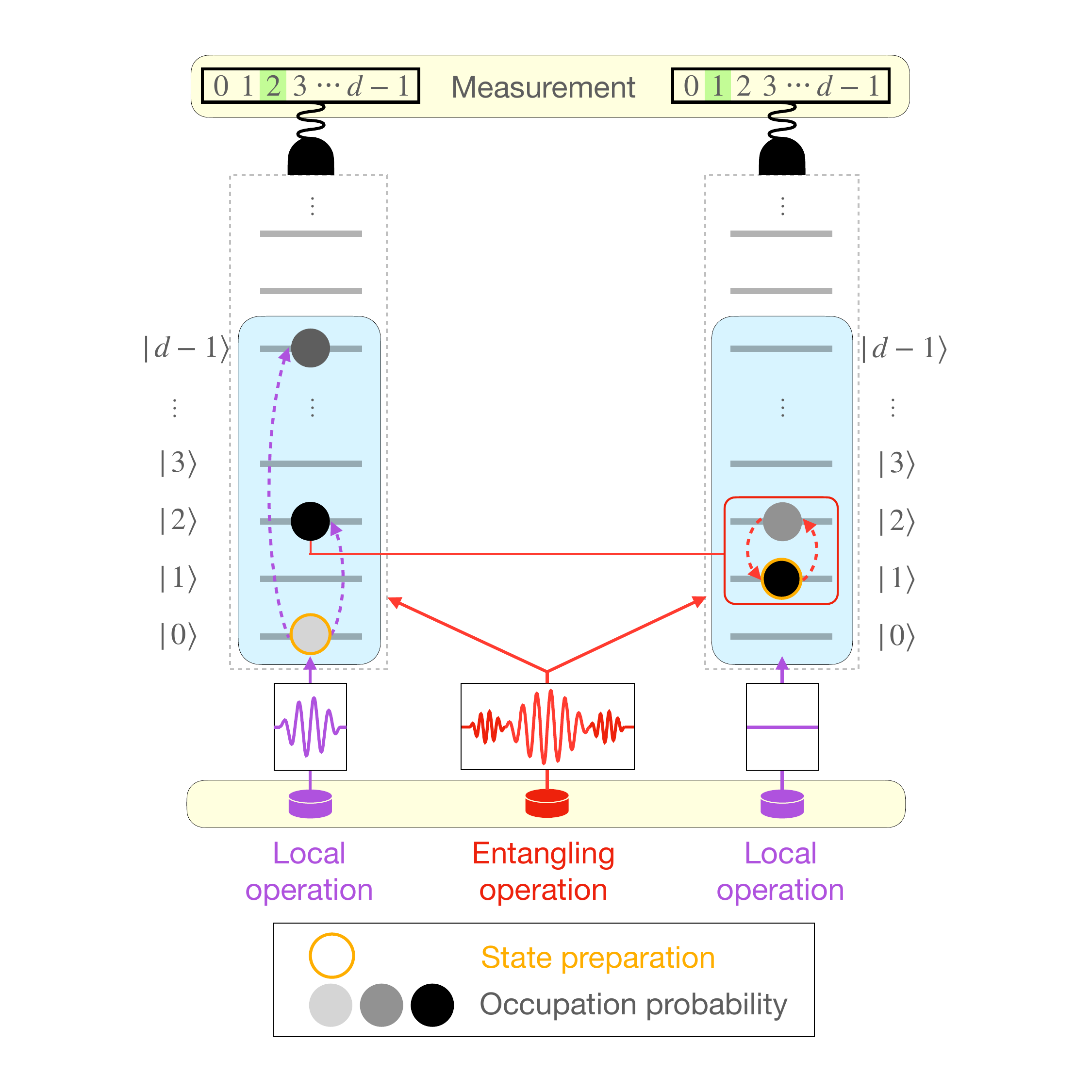}
    \caption{{\bf Qudit computing}. Within each physical information carrier (gray dashed box), there is a logical qudit (blue shade) with basis states (gray) $\ket{0}$ to $\ket{d-1}$, the span of which can be used for a computational task provided the qudit DiVincenzo criteria are satisfied (see Box~I).
    In this simple example, after state preparation within each qudit, a single-qudit gate and a two-qudit entangling gate generated by pulses from the control panel are applied to the qudits. Here the single-qudit gate drives the transitions $\ket{0}\leftrightarrow\ket{2}$ and $\ket{0}\leftrightarrow\ket{d-1}$ in the left qudit, while the two-qudit entangling gate drives the transition $\ket{1}\leftrightarrow\ket{2}$ in the right qudit conditioned on the left qudit being in $\ket{2}$. The final probability to find the qudit in each state is shown in grayscale. Both qudits are then measured, showing the outcomes $2$ and $1$, respectively.}
    \label{fig:intro}
\end{figure}

Since the inception of quantum computing~\cite{feynman_simulating_1982,deutsch_quantum_1985,feynman_quantum_1986}, quantum binary digits (qubits) are employed as convenient representations of quantum information.
Pedagogically, bits are convenient as the smallest radix, or base, that requires logarithmic space to store a number.
By extension,
dits, corresponding to higher radix (base~$d$),
can also be used to store a number:
boolean logic for bits is then replaced by higher-radix logic.
Under suitable conditions, dits can offer a more efficient information encoding.
In particular, quantum dits, or qudits, provide access to a larger usable Hilbert space for a fixed number of physical computing units (see Fig.~\ref{fig:intro})---much like a squeezer extracting more juice from an orange than just squeezing by hand---and thus potentially offer a faster route to impactful quantum computing.

The noisy intermediate-scale quantum (NISQ) paradigm has dominated efforts over the past decade~\cite{preskill_quantum_2018}.
However, a consensus is growing that fault tolerance is required to scale up the size, i.e., the dimension of useful Hilbert space (exponential in the number of qudits), and the depth, meaning the length of the quantum circuit in terms of the number of quantum logical operations that can be performed~\cite{eisert_mind_2025}.
Some refer to the current state of quantum computer technology as being `emerging fault tolerance'~\cite{katabarwa_early_2024}.
Experimental progress in preserving coherence across many quantum levels and realising high-fidelity qudit gates have reshaped this landscape, offering a path to scale to larger useful Hilbert space dimensions with fewer resources. These advances have resulted in a new wave of experiments demonstrating basic quantum computing protocols using qudits.

Quantum simulation, promoted early in the field by Feynman~\cite{feynman_simulating_1982}, can be thought of as a subset of quantum computation: it aims to study the properties of a specific physical system that is hard to simulate classically.
This can be accomplished either by evolving a controllable quantum system with an engineered Hamiltonian, or by performing a sequence of certain quantum logical operations. 
Conceptually, quantum simulation can be associated with quantum-state generation~\cite{aharonov_adiabatic_2007}
followed by measurement sampling,
under the principle of exploiting what nature provides. 
Using qudits rather than being restricted to qubits fits well into this framework and provides a new kind of ``pixel'' for quantum simulation, and recent experimental demonstrations on qudit platforms highlight the promise of leveraging multiple levels.

\begin{figure}[ht]
    \centering
    \includegraphics[width=0.55\linewidth]{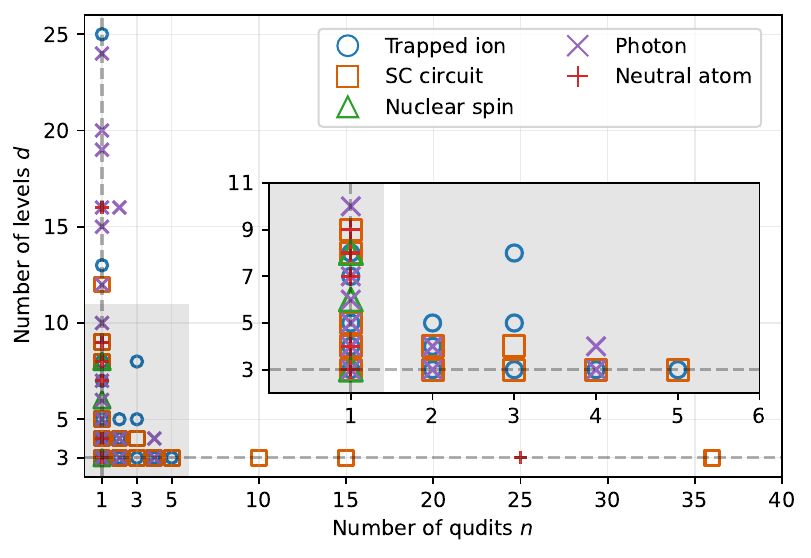}
    \caption{{\bf Current landscape of qudit experiments} (not an exhaustive list). Each marker at $(d,n)$ corresponds to an experiment that demonstrates arbitrary control of a single $d$-dimensional qudit, see Box~I,
    with $n$ being the largest number of qudits shown to participate in a non-disjoint computation or simulation with entangling interactions~\cite{footnote_photon_control} (note that $n$ is not the register size).
    Distinct symbols indicate different experimental platforms.
    Dashed lines indicate the special cases $n=1$ (single qudit) and $d=3$. The inset provides a magnified view of the shaded region in the lower-left corner, with $n=1$ separated from cases with entangling gates by a white strip.
    }
    \label{fig:d_N_fig}
\end{figure}

In recent years,
experimental quantum computing with qudits has rapidly advanced over diverse physical platforms, including trapped ions, superconducting circuits, solid-state nuclear spins, photonics and neutral atoms. Fig.~\ref{fig:d_N_fig} depicts the experimentally accessible regime in terms of the controllable qudit dimension $d$ and the number of qudits $n$ in a register, highlighting both current capabilities and the scope for further innovation. Experimental advances are being accelerated by developments in multi-level state preparation and measurement and related quantum control. These capabilities enable more quantum computational power to be drawn from systems of fixed size and depth by making use of more levels.

The remainder of this article is organised as follows: in Sec.~\ref{sec:control}, we discuss the experimental advances in state preparation, measurement, and control for qudit systems on various quantum computing platforms; Fig.~\ref{fig:intro} provides a schematic overview of these processes. 
We review recent experimental milestones in quantum simulation and quantum computing with qudits Sec.~\ref{sec:application}, along with demonstrations of error suppression strategies. Finally, we conclude by discussing challenges and opportunities for qudit quantum computing in Sec.~\ref{sec:outlook}.

\section{Advances in qudit platform development}\label{sec:control}
Qudits have now been realised across a wide range of quantum computing platforms that satisfy the DiVincenzo-like criteria described in Box~I. These include systems where multi-level structures arise naturally---such as trapped ions, solid-state nuclear spins, neutral atoms, and photons---as well as synthetic implementations in superconducting devices. The achievable levels of state preparation, measurement, and control (see Fig.~\ref{fig:intro}) vary across platforms, reflecting their distinct advantages and technical challenges.
In this section, we review the state of the art in state preparation and measurement (SPAM) and coherent control in the platforms that currently lead experimental qudit implementations; a summary is found in Table~\ref{tab:platforms}.
\vspace{-2em}

\subsection{Trapped ions}\label{subsec:trappedions}
A trapped-ion platform uses static and radio-frequency electric fields (a Paul trap) to confine individual ions in vacuum, where long-lived internal states are coherently controlled with lasers and entangled via shared motional modes. In current qubit processors, the $|0\rangle$ and $|1\rangle$ states are encoded in internal electronic levels (typically hyperfine or Zeeman sublevels). 
However, the multi-level electronic structure in ions can naturally support higher-dimensional Hilbert spaces for qudits: the anharmonic spectrum enables spectrally resolved transitions among many long-lived states, allowing qudit encoding within a single ion. Increasing the qudit dimension $d$ available for information processing introduces practical constraints,
including spectral crowding, variations in transition strengths, greater magnetic-field sensitivity, and longer state-discrimination sequences. Accordingly, trapped-ion qudit research has progressed along two directions: realising large-$d$ single-ion qudits to probe single-qudit SPAM and control limits, and scaling toward universal multi-ion processors with entangling gate fidelities approaching those of qubit systems.\\

{\it \noindent SPAM---}
In trapped ions, preparing an arbitrary basis state requires selective population transfer among closely spaced sublevels while suppressing off-resonant excitation and leakage. For large-$d$ encodings, such as the 25-level $^{137}\mathrm{Ba}^+$ qudit, narrow-band optical pumping schemes~\cite{low_quantum_2025} iteratively remove population from unwanted states using four laser wavelengths to isolate a target level. This approach uses nonlinear Zeeman shifts in the $D_{5/2}$ manifold to spectrally separate transitions, enabling high selectivity at the cost of longer preparation times and increased calibration complexity. For moderate dimensions (for example, seven levels in $^{40}\mathrm{Ca}^+$), simpler optical pumping followed by a single coherent pulse per state is typically sufficient, as spectral crowding and transition-strength inhomogeneity remain manageable~\cite{ringbauer_universal_2022, nikolaeva_scalable_2025,leupold_sustained_2018}.

\begin{FullWidthTwoColBox}\label{box:criteria}
\begin{center}
    {\large \bf Box I. DiVincenzo criteria adapted for qudit systems}
\end{center}

\begin{minipage}[t]{1\linewidth}

A qudit is a quantum system whose state resides in a~$d$-dimensional Hilbert space (Fig.~\ref{fig:intro}). Unlike a classical dit, which occupies a single value between 0 and $d-1$, a qudit can exist in an arbitrary coherent superposition of all~$d$ basis states, described by a complex vector of dimension~$d$. The $d$-ary quantum logic has additional mathematical structure that offers challenges and opportunities beyond the binary realm. In this box, we present a systematic definition of qudits in the context of quantum computing by extending DiVincenzo's criteria~\cite{divincenzo_topics_1997}, originally developed to identify viable qubit platforms for computing, to $d$-level systems~\cite{low_practical_2020}. We also assess the degree to which existing experimental platforms satisfy these requirements (see Table~\ref{tab:platforms}).

\begin{center}
\label{tab:criteria-comparison}
{\setlength{\tabcolsep}{4pt}
\renewcommand{\arraystretch}{1.3}
\setlength{\extrarowheight}{1.4pt}
\rowcolors{1}{gray!12}{white}
\begin{tabular}{|L{3.5cm}|L{2.3cm}|L{2.2cm}|L{2.4cm}|L{2.2cm}|L{2.6cm}|}
\hline
\textbf{Criterion} & \textbf{Trapped ions} & \textbf{SC circuits} & \textbf{Solid-state nuclear spins} & \textbf{Photons} & \textbf{Neutral atoms} \\
\hline \hline
1a. Well-characterised $d$-level systems &
{\large \checkmark}  & {\large \checkmark}  & {\large \checkmark}  & {\large \checkmark}  & {\large \checkmark} \\
\hline
1b. Scalability
 & Roadmap in \cite{low_practical_2020}  & Roadmap under development & Roadmap under development & Roadmap in \cite{malik_highdimensional_2026} & Roadmap under development\\
\hline
2. Initialisation &
{\large \checkmark}  & {\large \checkmark}  & {\large \checkmark}  & {\large \checkmark}  & {\large \checkmark} \\
\hline
3. Coherence &
{\large \checkmark}  & {\large \checkmark}  & {\large \checkmark}  & {\large \checkmark}  & {\large \checkmark}  \\
\hline
4a. Single-qudit arbitrary $\mathrm{SU}(d)$ gates &
{\large \checkmark}  & {\large \checkmark}  & {\large \checkmark}  & {\large \checkmark}  & {\large \checkmark} \\
\hline
4b. Two-qudit entangling gates &
{\large \checkmark}  & {\large \checkmark}  & Not yet demonstrated  & {\large \checkmark} &  Entangling interaction \\
\hline
5. Measurement &
{\large \checkmark}  & {\large \checkmark}  & {\large \checkmark}  & {\large \checkmark}  & {\large \checkmark}  \\
\hline 
\end{tabular}}
\end{center}
{
TABLE. {\bf Extended DiVincenzo criteria for qudits.}
{\large \checkmark} indicates an experimentally criterion for that platform.\\}

The criteria column in the table is to be interpreted as follows:
\begin{enumerate}
    \item \textbf{Scalable, well-characterised $d$-level systems:} The $d$ basis states must be well-defined, individually addressable, and sufficiently isolated (1a). 
    The information processing device should be scalable, accommodating a large number of such systems to perform tasks of practical utility (1b).
    \item \textbf{Initialisation:} Qudits are first prepared in a reference state that serves as a natural starting point of a computation.
    \item \textbf{Coherence:} Once initialised, the decoherence time for qudits across all $d$ levels should significantly exceed typical gate time to allow for performing a computational task involving execution of multiple gates.
    \item \textbf{Universality:} A universal qudit quantum computer should be able to efficiently approximate any unitary operator acting on the full Hilbert space to arbitrary precision.
    Operationally, this universality is typically decomposed into two ingredients: arbitrary single-qudit gates (4a) and at least one class of entangling two-qudit gates (4b) (see Box~II).
    \item \textbf{Measurement:} A qudit processor requires a measurement apparatus that can distinguish all~$d$ basis states.
\end{enumerate}

\end{minipage}
\hfill

\end{FullWidthTwoColBox}

Measurement of a trapped-ion qudit is usually performed via a sequential shelving and fluorescence readout scheme~\cite{ringbauer_universal_2022,low_control_2025,nikolaeva_efficient_2024}, with time cost scaling linearly in dimension. Fluorescence on the $S_{1/2}\rightarrow P_{1/2}$ transition identifies the $\ket{0}$ state as bright; if dark, selective transfer pulses map metastable $\ket{l\neq 0}$ states to the $S$ manifold, where subsequent detection confirms occupation. Heralded schemes that discard leakage events has achieved SPAM fidelities above $99\%$ for up to $25$ levels~\cite{low_quantum_2025}. Measurement can be further optimised by rearranging encoded states prior to readout~\cite{meth_simulating_2025}.
A major challenge in SPAM arises from sensitivity to magnetic-field fluctuations. Dressed qudit states have been proposed as a means to suppress magnetic-field noise~\cite{randall_efficient_2015}. Additional error sources include motional heating and dephasing, scattering-induced heating in larger ion chains, and crosstalk errors, all of which also affect qubit-based trapped-ion systems. In qudit encodings, these effects are often amplified by the larger Hilbert space.

\begin{FullWidthTwoColBox}\label{box:qudit-gates}
\begin{center}
    {\large \bf Box II. Strategies for qudit gate design}
\end{center}

The enlarged Hilbert space of qudits provides greater flexibility for quantum gate design.
Universal quantum computation in qudit platforms requires a universal qudit gate set:
a sufficient choice is the combination of arbitrary
single-qudit SU$(d)$
rotations and at least one two-qudit entangling gate. This box lists a few experimentally demonstrated strategies for realising such operations (see Sec.~\ref{sec:control} for information on concrete platforms). We note here that SU($d$) control is used as a shortened notation for arbitrary single-qudit SU$(d)/$U$(d-1)$ rotation and recognise that the former has $d^2-1-(d-1)^2=2(d-1)$ parameters for control and the latter has $d^2-1$ control parameters~\cite{yu_schrodinger_2025}---with arbitrary control achievable by concatenating a universal set of qudit logical gates.\\

\begin{minipage}[t]{0.48\linewidth}
\subsection*{Single-qudit gates}
Here we outline several examples for achieving arbitrary SU($d$) control.

\begin{itemize}[leftmargin=8pt]
    \item SU$(d)$ rotations can be constructed from $\mathcal{O}(d^2)$ SU$(2)$ operations acting on two-level subspaces of the~$d$-dimensional Hilbert space~\cite{reck_experimental_1994,rowe_representations_1999,clements_optimal_2016,deguise_simple_2018}. In practice, selection rules restrict these operations to directly connected state pairs.\\

    \includegraphics[width=1.0\linewidth]{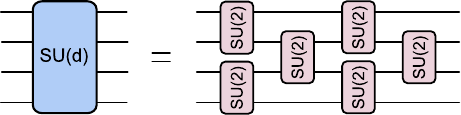}

    \item Quantum optimal control can be used to engineer pulse shapes that implements a generic SU$(d)$ rotation~\cite{walmsley_quantum_2003}. The control fields are parametrised—for example by amplitudes, phases, and durations—and optimised via a fidelity-based cost function to approximate the target unitary operator. The optimisation cost increases with qudit dimension $d$. Optimal control can also be used for generating two-qudit unitaries~\cite{omanakuttan_qudit_2023}.\\ 

    \includegraphics[width=1.0\linewidth]{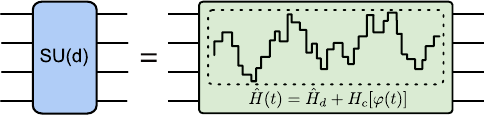}

    \item An arbitrary qudit unitary gate can also be synthesised by interleaving $N$ spin-displacement operators with $N+1$ diagonal phase rotations---selective number-dependent arbitrary phase (SNAP) gates---for $N\sim \mathcal{O}(d)$~\cite{champion_multifrequency_2025,roy_synthetic_2025,yu_schrodinger_2025,shi_efficient_2026}. Here, a spin-displacement gate (also commonly known as a covariant SU$(2)$ rotation) realises a rotation of an effective spin of size $j$, whose $d=2j+1$ levels are identified with the qudit levels.\\ 

    \includegraphics[width=1.0\linewidth]{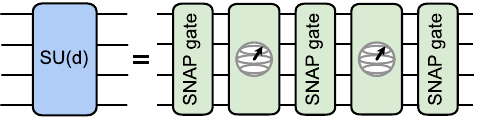}
    
\end{itemize}

\end{minipage}
\hfill
\begin{minipage}[t]{0.48\linewidth}

\subsection*{Two-qudit gates}
The presence of additional levels in a qudit enlarges the space of nonlocal gates that generates a richer variety of entanglement structure than is possible in the qubit case. Here, we provide some examples of qudit entangling gates.

A natural qudit entangling gate is the two-qubit controlled-rotation gate (such as the CNOT gate) restricted to a pair of target states (Gate~I in the figure below)~\cite{meth_simulating_2025,blok_quantum_2021}. This approach enables established two-qubit control techniques to be adapted to qudit architectures, with the added challenge of removing unwanted phases on spectator states.

Another possibility is the M\o lmer--S\o rensen (MS) type two-qubit rotation gate acting on two chosen states of each participating qudit (Gate~II in the figure)~\cite{ringbauer_universal_2022,nguyen_empowering_2024}. Using these qubit-type gates, one can also perform operations that involve multiple qudit states, e.g., a CSUM gate (a generalisation of the qubit CNOT gate to qudits) defined as CSUM $\ket{j, k} \rightarrow \ket{j,j\oplus k}$, where $\oplus$ denotes the sum operation modulo $d$~\cite{imany_highdimensional_2019}.\\

\includegraphics[width=1.0\linewidth]{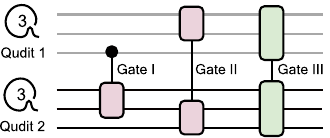}

~\\A further class of entangling operations comprises phase gates that act across all levels of the participating qudits (Gate~III in the figure), i.e., $G\ket{j,k}\rightarrow e^{i\theta_{j,k}}\ket{j,k}$~\cite{hrmo_native_2023,blok_quantum_2021,reimer_highdimensional_2019}.\\

Different type of entangling gates allow for different entanglement structures in the qudit systems. This flexibility enables the construction of circuits tailored to specific problem instances, with particular relevance for quantum simulation~\cite{zhang_synthesizing_2022,meth_simulating_2025}. Depending on the target interaction, gates can be chosen to act selectively on particular qudit states or, alternatively, operations that couple the full multi-level structure of the qudits.

\hfill
\end{minipage}

\end{FullWidthTwoColBox}

{\it \noindent Control---} 
A sufficient condition for universal qudit computation is the ability to implement arbitrary single-qudit rotations together with at least one entangling two-qudit gate (see Box~II for different strategies to implement these gates). A common approach to single-qudit control decomposes SU$(d)$ operations into sequences of SU$(2)$ rotations, each realised by resonantly driving a selected transition between sublevels (see Box~II). While compatible with existing qubit control techniques, this method faces a key challenge: off-resonant driving induces differential AC Stark (or Zeeman) shifts across all spectator levels. Unlike qubits, where only one relative phase is tracked, qudits accumulate multiple independent phases, requiring multi-chromatic compensation fields with up to $(d-1)$ tones, increasing calibration overheads~\cite{ringbauer_universal_2022}.

An alternative approach implements qudit rotations using simultaneous multi-tone control. In $^{137}\mathrm{Ba}^+$ experiments (up to $d=8$), a subset of internal states is chosen to ensure strong magnetic-dipole coupling, reduced magnetic-field sensitivity, and spectral isolation from leakage channels~\cite{shi_efficient_2026}. With an optimised quantisation field, multiple phase-coherent radio-frequency tones are applied simultaneously, generating collective spin–displacement operations (see Box~II) acting on all levels of the qudit. Properly engineered amplitudes and phases yield global SU$(d)$ rotations with circuit depth scaling linearly in $d$, avoiding long sequences of two-level gates. Similar optimal control techniques for pulse shape design has also been used to generate arbitrary single-qudit rotation with $>99\%$ fidelity~\cite{li_experimental_2022}.

Two-qudit entanglement has been realised in $^{40}\mathrm{Ca}^+$ and $^{171}\mathrm{Yb}^+$ systems using two distinct approaches. The first approach is to use M\o lmer--S\o rensen (MS) interactions~\cite{molmer_multiparticle_1999} restricted to two sublevels of each qudit to implement controlled-exchange or controlled-increment operations~\cite{aksenov_realizing_2023,ringbauer_universal_2022}. Despite additional overhead and noise compared to qubits, MS gates have been shown to achieve fidelities up to $99\%$ for trapped-ion qutrits~\cite{john_nonabelian_2026}. 

A second approach realises an entangling geometric phase gate. By generalising light-shift interactions and interleaving cyclic basis permutations, geometric phases are symmetrised across all levels and differential shifts are suppressed, enabling high-dimensional entanglement using a single gate~\cite{hrmo_native_2023}. Experiments demonstrate entanglement up to $d=5$ with higher fidelities than the MS gate---primarily limited by noise sources like Rabi fluctuations, motional decoherence and slow phase drifts.

Sideband-based controlled-rotation (Cirac--Zoller-type) gates~\cite{cirac_quantum_1995} provide another route to two-qudit interactions~\cite{meth_simulating_2025}, but require higher laser power due to Lamb--Dicke scaling of blue-sideband transitions. This induces kHz-scale AC Stark shifts that must be compensated with auxiliary off-resonant beams and software frame updates for spectator transitions; residual phase errors remain a dominant noise source.

\begin{table}
    \centering
    {\setlength{\tabcolsep}{4pt}
\renewcommand{\arraystretch}{1.3}
\setlength{\extrarowheight}{1.3pt}
    \rowcolors{1}{gray!12}{white}
\begin{tabular}{|
L{2.1cm}|
L{2.8cm}|
L{2.8cm}|
L{2.8cm}|
L{2.8cm}|
L{2.8cm}|}
        \hline
        \textbf{Platform} & \textbf{Trapped ions} & \textbf{SC circuits} & \textbf{Solid-state nuclear spins} & \textbf{Photons} & \textbf{Neutral atoms} \\
        \hline
        \vspace{-8em} Schematic  & \includegraphics[width=\linewidth]{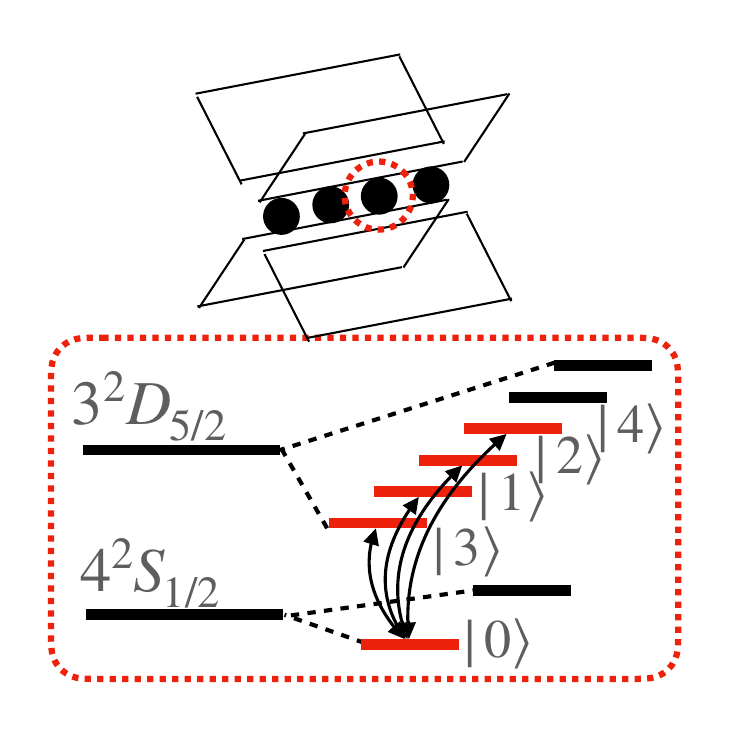} & \includegraphics[width=\linewidth]{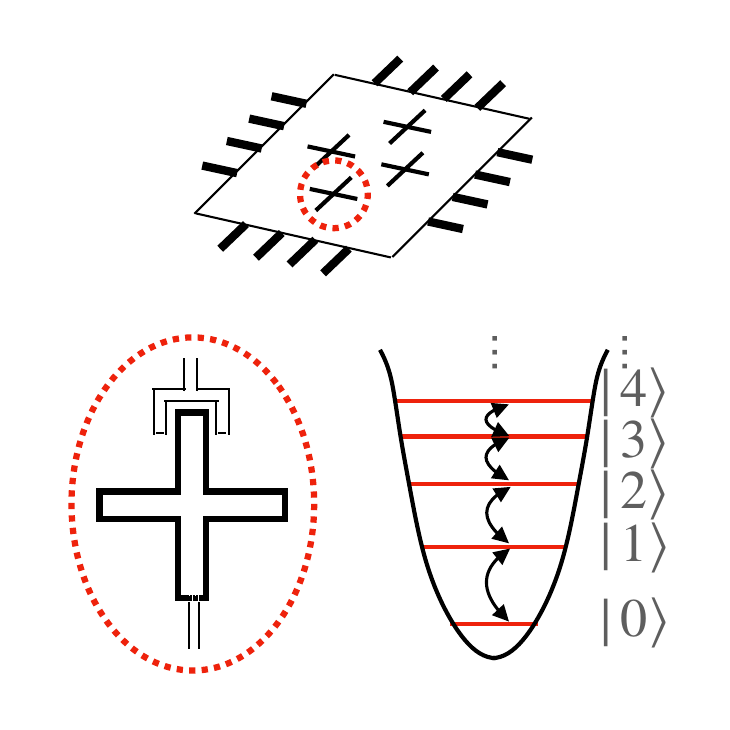} & \includegraphics[width=\linewidth]{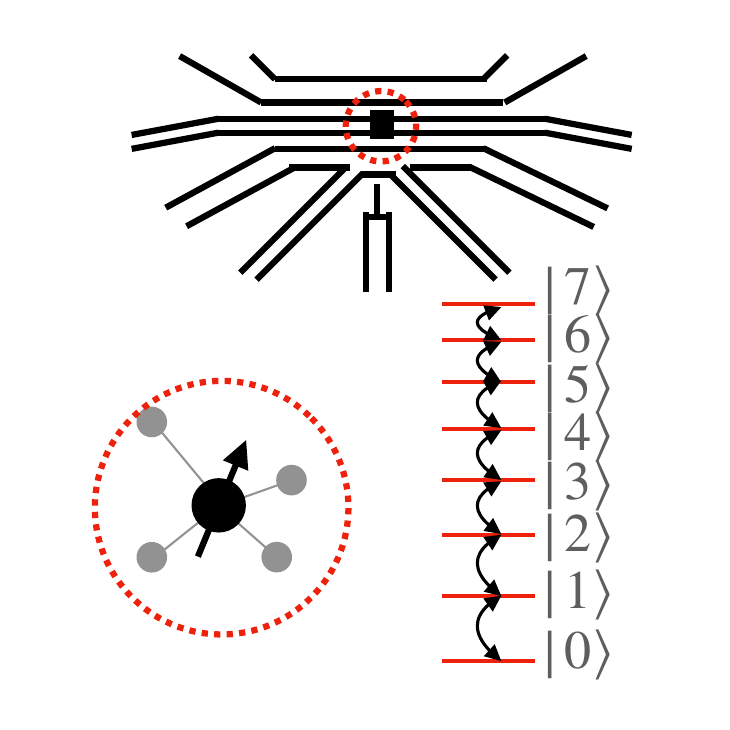} & \includegraphics[width=\linewidth]{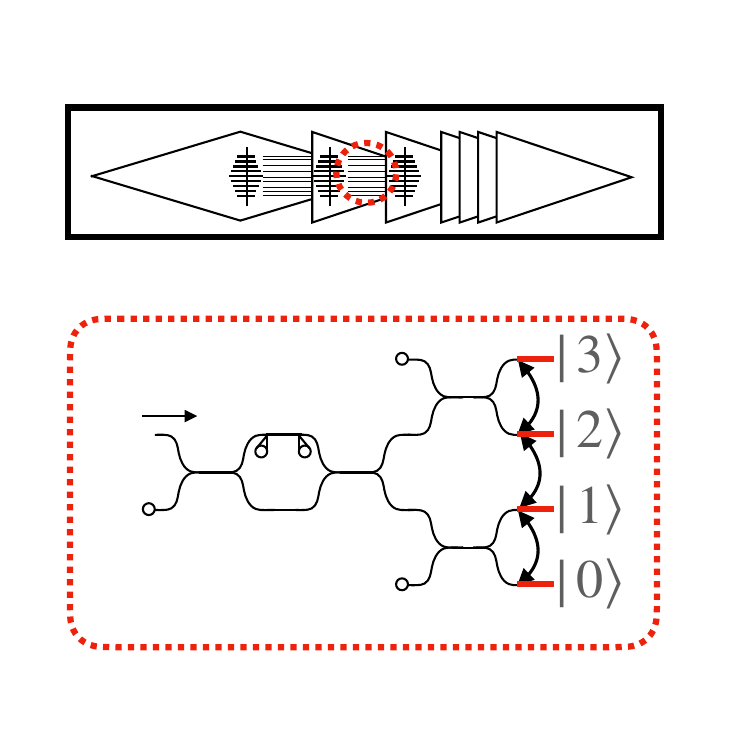} & \includegraphics[width=\linewidth]{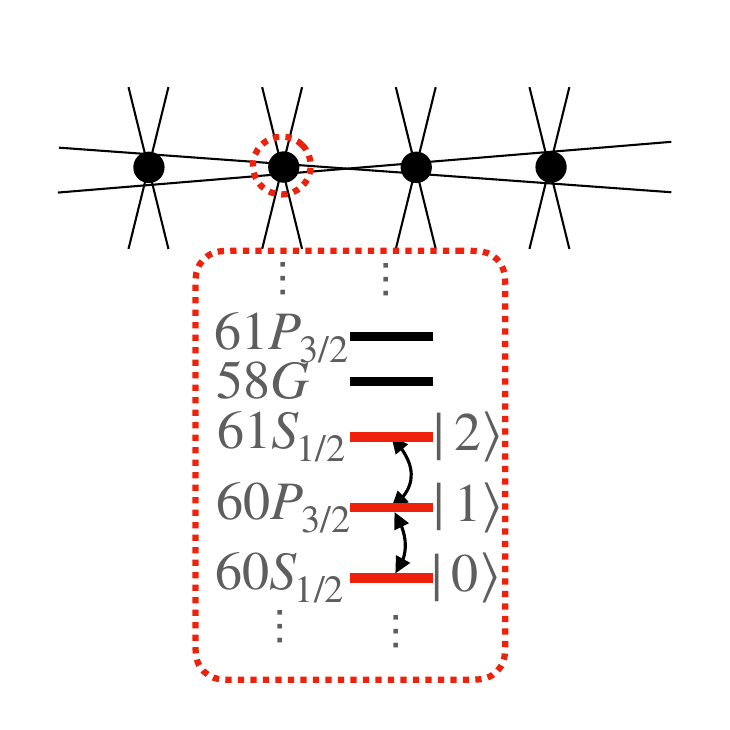}\\
        \hline
        Information carrier & Ions confined by electromagnetic fields & Electrical circuits with superconducting elements & Nuclear spins in solid-state systems & Photons in bulk or integrated optical systems & Neutral atoms in vapour cells, optical lattices or optical tweezer arrays \\
        \hline
        Qudit basis states & Electronic states of individual ions & Energy eigenstates of circuit units & Nuclear states of individual nuclei & Various, e.g. spatial modes, time, frequency, polarisation, or combinations thereof & Electronic and nuclear states of individual atoms \\
        \hline
        Largest~$d$ with arbitrary SU$(d)$ gate & $25$~\cite{low_quantum_2025} & $12$~\cite{wang_high_2025} &  $8\times 2$~\cite{fernandezdefuentes_navigating_2024} & $24$~\cite{barzaghi_lowloss_2025}  & $16$~\cite{smith_quantum_2013} \\
        \hline
        Largest~$d$ with arbitrary SU$(d)$ gate and entangling gates & $8$~\cite{john_nonabelian_2026}
        & $4$~\cite{nguyen_empowering_2024} & -- & $4$~\cite{chi_programmable_2022,liu_heralded_2026}  & $3$~\cite{qiao_realization_2025} \\
        \hline
        SPAM and control & Laser & Microwave and magnetic field & Microwave, magnetic field and electric field  & Various & Microwave, magnetic field and laser \\
        \hline
        Highest entangling gate fidelity & $98.7\% \ (d=3)$~\cite{hrmo_native_2023} & $97.3\% \ (d=3)$~\cite{goss_highfidelity_2022} & -- & $95.2\% \ (d=4)$ (probabilistic)~\cite{chi_programmable_2022} & Entangling interactions~\cite{qiao_realization_2025} \\
        \hline
        Qudit connectivity & All-to-all & Neighbouring & -- & Various & Finite-range \\
        \hline
    \end{tabular}
    }
    \caption{{\bf Overview of the state-of-the-art experimental qudit platforms focused on in this article.} Different qudit encoding schemes exist for photons and neutral atoms (see Sec.~\ref{sec:control}), and the schematics of path-encoded photonic qudits and Rydberg-atom qudits are shown as concrete examples. In \cite{fernandezdefuentes_navigating_2024}, an eight-dimensional nuclear-spin qudit is coupled to an electronic-spin qubit, making the total controllable dimension $8\times 2$.}
    \label{tab:platforms}
\end{table}

\subsection{Superconducting circuits}\label{subsec:SC_circuits}

On the superconducting platform, quantum information is stored and processed in nanofabricated macroscopic superconducting circuits. These highly customisable circuits are often configured such that their energy spectra resemble atomic ones; in such ``artificial atoms'', information is typically encoded in the lowest energy levels, although exceptions exist when low-energy auxiliary levels are required, e.g., for quantum simulation~\cite{tan_topological_2018,tan_experimental_2021} or error correction purposes~\cite{kubica_erasure_2023,liu_converting_2026}.

Most superconducting qudit experiments have employed transmons~\cite{koch_chargeinsensitive_2007, schreier_suppressing_2008}, a simple and scalable circuit design adopted by major players in the industry such as IBM~\cite{galda_implementing_2021,cervera-lierta_experimental_2022,kehrer_improving_2024,iiyama_quditgeneralization_2024,kumaran_transmon_2025} and Google~\cite{ticea_observation_2025}. In a transmon circuit, a Josephson junction---two superconducting electrodes separated by an insulating layer---is shunted by a large capacitance. When the Josephson energy $E_J$ far exceeds the capacitance charging energy $E_C$, the transmon becomes insensitive to charge noise (the main cause of dephasing error), and the spacing between adjacent energy levels slowly decreases as energy increases, ensuring that all transitions between adjacent levels are individually addressable by microwave pulses.

Transmons are popularly operated as qubits using their two lowest-energy eigenstates, but it is more challenging to operate them as qudits by including higher-energy states in the computational basis~\cite{blok_quantum_2021}. Higher-energy states have larger energy relaxation rates and dephasing rates~\cite{peterer_coherence_2015}; although increasing $E_J/E_C$ suppresses dephasing and accommodates more energy levels for encoding, it also makes different transitions less well-resolved, such that fast control pulses introduce more coherent error. Therefore, $E_J/E_C$ must be chosen carefully depending on the desired qudit dimension $d$: typically $\sim 50$ for strictly two-level experiments, $\sim 70$ for qutrits, and has been made as large as $325$ accommodating up to $12$ levels~\cite{wang_high_2025} (with control demonstrated up to $d=10$). Moreover, both classical and quantum crosstalk between qudits become more difficult to address as $d$ increases. Classical crosstalk arises from the spectral crowding of the transition frequencies for individual transmons, and can be mitigated by a combination of optimised device design, fabrication processes, and careful pulse engineering, whereas quantum crosstalk due to the cross-Kerr coupling can be mitigated by dynamical decoupling and using coupler transmons (see below).

In addition to transmons, there exist other circuit designs such as phase qudits~\cite{neeley_emulation_2009,svetitsky_hidden_2014} and flux qudits~\cite{yurtalan_implementation_2020,kononenko_characterization_2021,yurtalan_characterization_2021}. Yet another approach encodes logical qudits in the large Hilbert space of harmonic oscillator modes of superconducting cavities~\cite{brock_quantum_2025, roy_synthetic_2025,kim_ultracoherent_2025}, which have long coherence times and simple error channels (photon loss). Coupling to nonlinear elements such as transmon ancillae allows the universal control of cavities. Nevertheless, fast, high-fidelity control must be balanced with protecting cavity coherence against ancilla errors.\\

{\it \noindent SPAM---} Arbitrary single-qudit state preparation in transmons is accomplished by applying microwave pulses to the ground state $|0\rangle$. The simplest way to prepare $|0\rangle$ is passivel waiting for equilibration with the cold environment, but this approach is limited by the transmon relaxation time $T_1$. Fast active reset protocols based on measurement or engineered dissipation, originally developed for transmon qubits, are also effective for resetting transmon qudits~\cite{mcewen_removing_2021, marques_allmicrowave_2023}.

Standard qudit readout is performed by coupling the transmon to a readout resonator. A coupling that is weak compared to the transmon-resonator frequency difference results in a ``dispersive shift''---a shift of the resonator frequency that depends on the transmon state and can be measured by a readout microwave pulse.  In this dispersive limit, the measurement is quantum non-demolition (QND): repeated measurements always reproduce the first measurement outcome. Although the number of states that can be resolved with a single readout frequency is limited, fast, high-fidelity single-shot readout is achievable for large $d$ by using multiple readout frequencies~\cite{chen_transmon_2023,champion_multifrequency_2025}: for instance, Ref.~\cite{wang_high_2025} achieves the single-shot readout of a $d=10$ transmon qudit using three spectrally distinct readout tones. Finally, qudit readout has proven a powerful diagnostic tool for studying measurement-induced leakage errors during standard qubit readout, which can involve multi-photon excitations to highly excited transmon states~\cite{wang_probing_2026}. \\

{\it \noindent Control---} Single-qudit $\mathrm{SU}(d)$ operations are usually implemented by sequences of two-level rotations realised through microwave control pulses, together with noiseless virtual phase gates that update the pulse phases on the software level~\cite{blok_quantum_2021, morvan_qutrit_2021,fischer_ancillafree_2022,liu_performing_2023} (see Box~II). 
Several transitions can be driven simultaneously, potentially further reducing the gate time and improving fidelity~\cite{kumar_stimulated_2016,danilin_experimental_2018,vepsalainen_superadiabatic_2019,dogra_experimental_2022}. If $d$ is not too large, such decompositions can be straightforward for simple gates such as generalised Pauli and Hadamard gates~\cite{neeley_emulation_2009, yurtalan_characterization_2021,dong_simulation_2021,yu_efficient_2025}. Hardware-specific optimal control techniques can be employed to directly implement arbitrary single-qudit gates at the pulse level without decomposing them into primitive gate sequences~\cite{zheng_optimal_2022,zheng_demonstration_2022,seifert_exploring_2023,cho_direct_2024}, although simple gates can also benefit from optimal control~\cite{wu_highfidelity_2020}. Alternatively, the decomposition into spin-displacement gates and virtual phase gates has also been demonstrated~\cite{roy_synthetic_2025,champion_multifrequency_2025}, with each spin-displacement gate implemented by driving all $d-1$ adjacent transitions simultaneously.

The qudit controlled-rotation gate (see Box~II) can be realised via the cross-resonance interaction for statically coupled transmons. This is commonly employed for implementing qubit CNOT gates and has been generalised to qutrits~\cite{blok_quantum_2021,galda_implementing_2021,cervera-lierta_experimental_2022,kumaran_transmon_2025}. When a drive pulse is applied to transmon 1 at the $|0\rangle$--$|1\rangle$ frequency of transmon 2, transmon 2 experiences a Rabi oscillation whose rate depends on the state of transmon 1. For an appropriate gate time, the cross-resonance gate can be made to realise a $|0\rangle$--$|1\rangle$ rotation in transmon 2 conditioned on transmon 1 being in $|1\rangle$.

The qudit MS gate (see Box~II) has been demonstrated between capacitively coupled fixed-frequency transmons. By driving both transmons at the same frequency that is detuned from the single-photon transitions $|j,k\rangle \to |j+1,k\rangle (|j,k+1\rangle)$, one can ``dress'' the $|j+1,k\rangle$ and $|j,k+1\rangle$ states, modifying their energies. A coherent two-photon hopping process between $|j,k\rangle$ and $|j+1,k+1\rangle$ is then realised through sequential virtual single-photon transitions mediated by the dressed states~\cite{nguyen_empowering_2024}. Similarly, one can effect a coherent $|11\rangle \leftrightarrow |02\rangle$ interaction by applying to transmon 2 a microwave pulse detuned from its $|1\rangle \leftrightarrow |2\rangle$ transition, such that the $|11\rangle$ and $|02\rangle$ states become degenerate~\cite{nguyen_programmable_2024}. Notably, it is also standard practice to use magnetic flux modulation in tunable transmons to implement the qubit CZ gate, where the $|11\rangle \leftrightarrow |02\rangle$ process creates a geometric phase in $|11\rangle$~\cite{dicarlo_demonstration_2009,chen_qubit_2014,caldwell_parametrically_2018}. Both $|11\rangle \leftrightarrow |22\rangle$ and $|11\rangle \leftrightarrow |02\rangle$ have been leveraged for the low-depth synthesis of multi-qubit Toffoli gates, with $|2\rangle$ used as built-in ancillae~\cite{fedorov_implementation_2012,chu_scalable_2023,nguyen_programmable_2024}.

The controlled-phase gate (see Box~II) has also been realised between transmon qudits through their cross-Kerr interaction, which describes the frequency shift of one transmon that depends on the other's state. The cross-Kerr interaction exists as a static interaction between capacitively coupled transmons~\cite{blok_quantum_2021}, but can also be dynamically generated as a differential AC Stark shift when both transmons are driven off-resonantly at the same frequency~\cite{goss_highfidelity_2022}. Alternatively, the cross-Kerr interaction can be engineered between two transmons by applying flux pulses to a third flux-tunable ``coupler'' transmon capacitively coupled to both transmons~\cite{luo_experimental_2023,roy_twoqutrit_2023}. The Stark-driven and tunable-coupler schemes achieve larger on-off ratios than the static scheme, and are thus more effective in suppressing the quantum crosstalk.

\subsection{Solid-state nuclear spins} \label{subsec:nuclear-spins}
Nuclear spins provide high-dimensional Hilbert spaces ($I > 1/2$) with long coherence times due to weak magnetic moments that suppress environmental coupling. Notable examples in solid-state devices include $I=7/2$ antimony donor in silicon ($d=8$)~\cite{fernandezdefuentes_navigating_2024}, $I=7/2$ vanadium in silicon carbide ($d=8$)~\cite{koller_strainenabled_2025}, $I=5/2$ manganese donor in zinc oxide crystal ($d=6$)~\cite{lim_demonstrating_2025}, and $I=3/2$ single molecular magnet $\mathrm{TbPc}_2$ ($d=4$)~\cite{godfrin_operating_2017,godfrin_generalized_2018}. Recent experiments demonstrate single-qudit gate fidelities exceeding 99.8\% in semiconductor donor systems~\cite{fernandezdefuentes_navigating_2024}.  Entangling gates between solid-state nuclear qudits have not been demonstrated so far; however, entanglement between nuclear qubits is well-studied~\cite{neumann_multipartite_2008,klotz_bipartite_2026}, and antimony-pair doping techniques have demonstrated a pathway to fabricating coupled nuclear qudit arrays~\cite{jakob_scalable_2024,adshead_isotopically_2026}.\\

{\it \noindent SPAM---} 
State preparation and readout require spectral resolution and selective addressing of all transitions achieved via anharmonicity in nuclear Zeeman levels. This anharmonicity arises from hyperfine interactions~\cite{lim_demonstrating_2025} or quadrupole coupling induced by electric field gradients, for example from local strain in the silicon lattice with antimony donors~\cite{fernandezdefuentes_navigating_2024}. Initialisation utilises either a readout of the nuclear spin followed by resonant $\pi$-pulses (nuclear magnetic resonance (NMR)~\cite{yu_schrodinger_2025} or nuclear electric resonance (NER)~\cite{asaad_coherent_2020}), or spin pumping by AC electric fields that modulate the hyperfine interaction and induce electron-nuclear flip-flop transitions in an electric dipole spin resonance (EDSR)~\cite{fernandezdefuentes_navigating_2024}.

In systems involving donor atoms in silicon, the nearly QND nuclear spin measurement leverages a hyperfine-coupled electron that minimally perturbs the nuclear state~\cite{vaartjes_certifying_2025,yu_schrodinger_2025,asaad_coherent_2020,fernandezdefuentes_navigating_2024}. A nuclear-state-dependent ESR pulse is swept across all transition frequencies of the qudit. A spin flip at a specific frequency, followed by spin-to-charge conversion detected with a single-electron transistor, identifies the qudit basis state and enables projective single-shot readout.

Initialisation and measurement in TbPc$_2$ are both achieved by sweeping a magnetic field across hyperfine anticrossings at millikelvin temperatures~\cite{godfrin_operating_2017,godfrin_generalized_2018}. Quantum tunnelling of magnetisation occurs at field values set by nuclear-state-dependent anticrossings between states with the same nuclear spin and opposite electron spins, producing distinct conductance signatures and enabling projective readout of the four-level qudit.\\

{\it \noindent Single-qudit control---} Coherent control relies on implementing Givens rotations involving two sublevels of a qudit. These are realised magnetically via NMR~\cite{yu_schrodinger_2025} or electrically via NER (which modulates the electric quadrupole interaction)~\cite{asaad_coherent_2020,fernandezdefuentes_navigating_2024} or EDSR~\cite{fernandezdefuentes_navigating_2024}. Rabi frequencies range from tens of Hz (electric driving) to tens of kHz (magnetic driving), while coherence times of 0.1--1~s enable thousands of operations. Gate fidelities above 99.3\% have been achieved for certain single-qudit operation and state preparation~\cite{fernandezdefuentes_navigating_2024,vaartjes_certifying_2025}. Electric quadrupole control in NER enables highly local, scalable addressing and integrability in nanoscale devices but is typically weaker, requiring engineered strain or large field gradients to achieve strong transitions. Conversely, magnetic (NMR) control is faster and robust but harder to localise.

In TbPc$_2$, three microwave NMR transitions are resolved by hyperfine and quadrupole splitting~\cite{godfrin_generalized_2018}. MHz-scale Rabi oscillations and millisecond coherence times enable $>10^3$ operations. Simultaneous multi-frequency pulses have been used to realise generalised Hadamard gates and a toy realisation of a quantum search algorithm within a single four-level qudit~\cite{godfrin_operating_2017}, demonstrating non-trivial computation without entangled qubits.

\subsection{Photons}\label{subsec:photons}
Photons offer a versatile platform for qudit encoding in multiple degrees of freedom (DoFs) that can also be combined to access larger Hilbert spaces without substantially increasing device footprint. Here we provide a brief overview of common encoding schemes with a focus on qudit quantum computing, and refer to other excellent reviews for further detail~\cite{forbes_quantum_2019,erhard_advances_2020,lu_frequencybin_2023,singh_photonic_2025,montaut_progress_2025,chang_recent_2025,goel_quantum_2025,malik_highdimensional_2026}. It is worth noting that pioneering advances in high-dimensional photonic entanglement generation have reached dimensions as large as $d \sim 1000$~\cite{chang_recent_2025}; even without full SU$(d)$ control, these systems can provide a powerful resource for quantum information processing~\cite{footnote_photon_control}.

In most experiments, photon are generated in pairs probabilistically via spontaneous parametric down-conversion or four-wave mixing, and detected using single-photon detectors with coincidence counting~\cite{eisaman_invited_2011}. The preparation and manipulation of photonic qudits depend strongly on the encoding scheme, which broadly falls into spatial and time–frequency categories, while polarisation alone is limited to $d=2$.\\

{\it Path encoding---} Spatial encoding includes path and transverse-mode encoding approaches. In path encoding, each basis state corresponds to a distinct optical path, enabling initialisation via coherent pumping of multiple sources and manipulation using beam splitters and phase shifters. This bottom-up approach supports arbitrary single-qudit unitaries and is compatible with integrated photonics~\cite{carolan_universal_2015}. Two-qudit entangling gates have been demonstrated probabilistically on path-encoded ququarts: the success probability of the multiqudit multi-value controlled-unitary gate scheme~\cite{chi_programmable_2022} scales as $1/d$ independent of the number of qudits, whereas that of the post-selected fusion gate~\cite{vigliar_errorprotected_2021} between two photons is $1/2$, independent of $d$. Although scaling to large $d$ is limited by the need for additional paths and quadratic growth in optical elements, integrated implementations have reached $d=24$~\cite{barzaghi_lowloss_2025}.\\

{\it Transverse spatial‑mode encoding---} The computational basis is encoded in spatially structured modes such as orbital angular momentum (OAM), pixel modes, multi-mode waveguides, or multi-core fibres~\cite{goel_quantum_2025}. Qudit operations are implemented using phase-only transformations (e.g., spatial light modulators) or mode-mixing elements such as multi-mode waveguides and scattering medium. Dove prisms and spiral phase plates can also perform single-qudit unitary transformation on OAM qudits. Heralded entangling gates for OAM ququarts have been demonstrated using polarisation as an auxiliary DoF, two auxiliary OAM qudits, and Bell-state measurements~\cite{liu_heralded_2026}, with success probability $1/8$ independent of $d$ under feedforward conditioned on the measurements.
Transverse spatial-mode encoding has also been used for generating a two-photon cluster state with four ququints~\cite{lib_resourceefficient_2024}.
While top-down approaches like this are robust to fabrication errors, OAM modes remain challenging for integration and transmission in standard single-mode fibres, although special-purpose fibres such as ring-core fibres have shown promise~\cite{liu_optical_2026}.\\

{\it Time–frequency encodings---} Photonic qudits can be encoded in temporal (time-bin) or spectral (frequency-bin), or hybrid time-frequency modes~\cite{lu_frequencybin_2023,singh_photonic_2025}. Time-bin qudits are generated using interferometers or by pumping with tailored pulse sequences and can be converted to path encoding for manipulation. Frequency-bin qudits are prepared via spectral filtering and controlled using electro-optic modulators and pulse shapers.  Deterministic unitaries controlled by the frequency DoF and acting on the time DoF have been demonstrated using a dense wavelength-division multiplexing filter to separate frequency bins (a controlled-increment gate)~\cite{imany_highdimensional_2019}, or a fibre Bragg grating array to implement frequency-dependent time delays (a controlled-phase gate or a CSUM gate, see Box II)~\cite{reimer_highdimensional_2019,imany_highdimensional_2019}. These techniques have enabled the realisation of a four-particle cluster state with $d=3$ and proof-of-principle measurement-based quantum computation with qutrits~\cite{reimer_highdimensional_2019}.
Time–frequency encodings are attractive because of their compatibility with integrated photonics, robustness to environmental noise, and efficient transmission in single-mode fibres. Their main limitation is the need for active or nonlinear components, which increases experimental complexity when compared with passive spatial schemes.

\subsection{Neutral atoms}\label{subsec:neutral-atoms}
Neutral atoms, like trapped ions, naturally provides a $d$-dimensional $(d>2)$ Hilbert space encoding through their electronic and nuclear-spin DoFs. In the case of electronic spins, qudits can be encoded in ground-state hyperfine manifolds or Rydberg states.\\

{\it Hyperfine encoding---} Hyperfine qudits are initialised via optical pumping into a stretched Zeeman state, followed by coherent control using magnetic, radio-frequency (RF), or microwave fields. Early work achieved near-universal control of the $F=3$ manifold of $^{133}\mathrm{Cs}$ by combining time-dependent magnetic fields with tensor AC Stark shifts, enabling arbitrary state mappings in a seven-dimensional space with yields $\sim0.9$~\cite{chaudhury_quantum_2007}. This was later extended to the full $16$-dimensional ground-state manifold ($F=3,4$) using optimal-control-designed phase-modulated RF and microwave fields, reaching fidelities above $99\%$~\cite{smith_quantum_2013}. Readout is performed via state-selective fluorescence to resolve population in the qudit levels. Qudit control performance is limited by magnetic-field noise, control-field inhomogeneity and optical scattering.\\

{\it Rydberg encoding---} Rydberg atoms, with highly excited valence electrons, exhibit strong, tunable long-range interactions~\cite{saffman_quantum_2010,adams_rydberg_2019}, which scale as $1/R^3$ at short range and $1/R^6$ at larger separations. Interaction strengths can be selected by choosing appropriate electronic states and engineered using off-resonant laser or microwave dressing. Qudits may be encoded directly in Rydberg levels~\cite{qiao_realization_2025}, or in ground and low-lying states coupled via Rydberg interactions~\cite{omanakuttan_qudit_2023,jia_architecture_2024,burshtein_robust_2026}. A qudit encoding scheme using collective Rydberg excitations of atomic ensembles have also been demonstrated, which is robust against atom loss~\cite{jiao_single_2025}.

State preparation involves loading atoms into optical tweezer arrays or lattices, followed by laser cooling and optical pumping. Population can then be coherently transferred to the desired qudit state using resonant excitation or adiabatic techniques such as STIRAP. Single-qudit gates are implemented by microwave~\cite{jiao_single_2025} or optical pulses,  while site-resolved light shifts allow selective detuning of individual atoms from globally applied fields~\cite{qiao_realization_2025}. Readout is typically performed by mapping Rydberg states to ground states and detecting via site-resolved fluorescence imaging, or by converting collective excitations to photonic modes~\cite{jiao_single_2025}. State-preparation fidelities approach $\sim95\%$ per site~\cite{qiao_realization_2025}, with limitations arising from finite Rydberg lifetimes, residual interactions, and Doppler dephasing. 

A qutrit-based simulation of interacting spins and hole states has been performed on the Rydberg platform using the native interactions between Rydberg atoms~\cite{qiao_realization_2025}. Qudit entangling gates have also been proposed based on the Rydberg blockade mechanism~\cite{gonzalez-cuadra_hardware_2022,zache_fermionqudit_2023, omanakuttan_qudit_2023,jia_architecture_2024, burshtein_robust_2026} where a single excitation shifts neighbouring atoms out of resonance. \\

{\it Nuclear-spin encoding—} Nuclear-spin qudits have been realised in neutral $^{87}\mathrm{Sr}$ atoms~\cite{krondorfer_singlequdit_2025,ahmed_coherent_2025}. Initialisation is achieved by optical pumping into a particular nuclear-spin state, followed by cooling and removal of residual Zeeman components. The resulting manifold provides a long-lived multi-level system for coherent control.

Control can be implemented using optical nuclear electric resonance (ONER), where electric-field gradients couple to the nuclear quadrupole moment, enabling state-selective Raman transitions~\cite{krondorfer_singlequdit_2025}. Simulations with realistic noise models predict amplitude-modulated optical fields can address all ten nuclear-spin transitions via ONER with $\pi$-pulse fidelities above $99.9\%$. Optical control enables fast (kHz-scale) manipulation combined with long coherence times, supporting scalable qudit registers in optical lattice architectures. Alternatively, coherent control can be achieved with two phase-coherent laser beams: one inducing tensor light shifts generating quadratic energy splitting between Zeeman states and the other driving Raman transitions between them~\cite{ahmed_coherent_2025}.\\

It is also possible to encode a qudit in the momentum-component states of a Bose--Einstein condensate (BEC) of neutral atoms; this has been demonstrated for $^{87}\mathrm{Rb}$~\cite{flament_unitary_2025}, where the BEC is initialised by adiabatic loading onto an optical lattice, controlled by adjusting the lattice position, and read out by abruptly switching off all traps and taking an absorption image after a certain time of flight.

\subsection{Other platforms}

Among other platforms exploring qudit quantum computing, vacancy centres and molecular systems have seen notable recent progress. In solid-state defects (NV in diamond, SiC divacancies) initialisation is dominated by optical pumping into a preferred electronic and nuclear basis state and readout by spin-dependent photoluminescence, while coherent control exploits spectrally selective microwave and RF drives, multi-tone pumping and rotating-frame Hamiltonian engineering to address adjacent $\Delta m=\pm 1$ and double-quantum transitions with sub-MHz spectral resolution~\cite{soltamov_excitation_2019,chen_synthetic_2022,fu_experimental_2022}. Molecular qudits exploit large intrinsic hyperfine coupling between the electronic and nuclear manifolds~\cite{chicco_controlled_2021,chicco_proofofconcept_2024,hussain_coherent_2018}, although the nuclear quadrupolar interaction need not be strong~\cite{gimeno_broadband_2021}. Targeted RF pulses realise high-contrast, leakage-suppressed Rabi flops and Hahn-echo times that can reach milliseconds to tens of milliseconds for nuclear transitions, while multiple-quantum coherences remain the one of the main bottlenecks \cite{chicco_proofofconcept_2024}. Finally, rotational states of ultracold polar molecules demonstrate a different regime: by combining magic-wavelength trapping and microwave control, coherent multi-level superpositions become accessible, opening metrological and simulation applications~\cite{hepworth_longlived_2025}.

\section{Applications}\label{sec:application}
The advances in high-fidelity control discussed in Sec.~\ref{sec:control} naturally raise the question of how these capabilities can be used to unlock the full potential of qudits. This question is now being actively explored across academia and industry, leading to  first proof-of-concept applications in both quantum simulation (Table~\ref{tab:simulation}) and universal quantum computation (Table~\ref{tab:algorithms}).
Classical simulation of quantum systems usually requires resources that grow exponentially with system size due to the Hilbert space dimension. Using quantum simulations, we can probe the properties of such quantum systems without incurring this exponential overhead, allowing us to reach a system size beyond the reach of classical computers.
Quantum simulation therefore provides a natural route to quantum advantage while advancing our understanding of physical systems. 
By contrast, quantum computation runs algorithms on quantum hardware to solve problems that need not originate from physical systems, with the goal of achieving computational speedups over classical methods. Achieving this requires universal control of the hardware. In this section, we survey recent progress in quantum simulation and universal quantum computation enabled by advances in qudit control.

\subsection{Quantum simulation}\label{subsec:simulation}
Multi-level DoFs arise in many physical systems, including high-energy physics, condensed matter physics, and quantum chemistry, and can be naturally mapped to qudits (see Table~\ref{tab:simulation}). Qudit encodings avoid non-local terms that appear when a $d$-dimensional DoF is mapped onto multiple qubits, thereby reducing circuit depth and entangling-gate counts.
The first proof-of-concept qudit simulations of lattice gauge theories have already demonstrated an entangling-gate count reduction by more than one order of magnitude~\cite{meth_simulating_2025,john_nonabelian_2026}: for instance, a circuit that has been implemented on the trapped-ion platform with three $d=8$ qudits and $67$ entangling gates would require $1088$ CNOT gates on a qubit processor.
Conversely, multiple two-level DoFs can be compactly embedded within a single qudit, resulting in shallower and more resource-efficient circuits. Both approaches allow simulations to scale to larger effective system sizes with a fixed experimental register size.\\

{\it Lattice gauge theories---} Gauge theories underpin much of our understanding of nature, describing phenomena ranging from quark confinement in high-energy physics to superconductivity in condensed matter physics. To go beyond the weak-coupling regime, gauge theories are discretised on a lattice, allowing the efficient classical computation of certain properties. In the most general case, however, classical algorithms become inadequate, necessitating the quantum simulation of lattice gauge theories. The infinite-dimensional representations of the gauge fields are usually truncated to a finite, preferably large dimension $d$ before being implemented on a quantum simulator; choosing $d>2$ yields a natural mapping onto a qudit. This has been demonstrated experimentally in a trapped-ion platform, where variational ground-state preparation and real-time dynamics of a \((2+1)\)-dimensional lattice quantum electrodynamics unit cell were realised~\cite{meth_simulating_2025}. The pure gauge theory was simulated using qudits with \(d=3\) and \(d=5\) in a three-ion register; in the theory with matter, the hybrid qubit--qudit circuit contained four matter qubits and one gauge qutrit, with matter--gauge interactions implemented via qubit--qudit entangling gates. Non-Abelian real-time string breaking dynamics in a pure-gauge $\mathrm{SU}(2)$ theory were also simulated using three trapped-ion qudits with $d=4$ and $d=8$~\cite{john_nonabelian_2026}. Several platforms allow such on-demand choice of the qudit dimension for each physical site, enabling interactions between subsystems of different dimensionality and offering flexibility in encoding strategies and circuit design. On superconducting hardware, the $(1+1)$-dimensional Abelian Higgs model has been simulated using two transmon qutrits~\cite{asaduzzaman_hybrid_2026}, with the gauge field Hilbert space truncated to $d=3$. In parallel with a digital protocol, the time evolution was also performed with a hybrid analog-digital protocol, where the target interaction between gauge fields were generated from the native cross-Kerr interaction by Floquet engineering and digital dynamical decoupling sequences.\\

{\it Higher spin models---} Multi-level systems also appear in spin models with $S>1/2$;
e.g., three levels of a spin-$1$ particle can be mapped onto a single qutrit instead of two qubits. The discovery of symmetry-protected topological phases in spin-$1$ chains has motivated qutrit-based quantum simulations of spin-$1$ systems~\cite{senko_realization_2015}, such as the preparation of the Affleck--Kennedy--Lieb--Tasaki state within the symmetry-protected Haldane phase~\cite{kumaran_transmon_2025,edmunds_symmetryprotected_2025}. Out-of-equilibrium, periodically driven (Floquet) spin-$1$ systems have also attracted interest: dynamical localisation has been observed in a chain of four trapped-ion qutrits~\cite{camacho_observing_2024}, and a $\mathbb{Z}_3$ discrete time crystal (DTC) has been probed in an array of $15$ superconducting qutrits, exhibiting qualitatively distinct behaviour from a DTC in the $\mathbb{Z}_2$ Ising model such as intrinsically chiral interactions~\cite{goss_qutrit_2026}.
In another high-spin example, the Lipkin-Meshkov-Glick model---a testbed model in many-body physics that can be represented by a large collective spin $S$ and host rich quantum critical phenomena---has been simulated for $S=4$ on a superconducting transmon qudit with $d=9$~\cite{champion_analog_2025}. Finally, higher spin models can also arise in chemical systems where unpaired localised electrons can often be modelled as effective local spins with \(S \geq 1/2\)~\cite{malrieu_magnetic_2014,maskara_programmable_2025}.\\

{\it Other multi-level systems---} Beyond purely spin models, qudits provide natural encoding choice in other bosonic models in condensed matter physics. A quantum antiferromagnet model of hard-core bosons has been simulated using Rydberg-atom qutrits encoding spin-up, spin-down and hole states~\cite{qiao_realization_2025}. Superconducting transmon qutrits have been used to simulate the disordered Bose-Hubbard model with at most two bosons per site, yielding insight on the transition between the Mott insulator and superfluid phases in the presence of disorder~\cite{ticea_observation_2025}. Qutrits have also been used to simulate neutrino oscillations with three neutrino flavours~\cite{spagnoli_collective_2025,nguyen_simulating_2023}, and to simulate quantum information scrambling in many-body physics~\cite{blok_quantum_2021}. Furthermore, qudits have been employed to simulate multi-level donor--acceptor dynamics in electron transfer pathways~\cite{sun_quantum_2023}, molecular chirality discrimination~\cite{liu_chiral_2025}, quantum walks~\cite{zhou_demonstration_2025}, and topological monopoles~\cite{tan_topological_2018,sugawa_second_2018,tan_experimental_2021,chen_synthetic_2022}, reflecting a natural fit of the qudits in the simulator to the target Hilbert space dimensions. \\

{\it Multi-qubit-to-qudit encoding and access to additional levels---} Experimental demonstrations of qudit utility also include multiple logical qubits encoded in a single qudit. A single $d=16$ qudit has been used to encode a four-qubit $\mathrm{LiH}$ Hamiltonian and perform variational ground-state preparation without entangling gates~\cite{kim_quditbased_2024}, demonstrating significant reductions in entangling resources. Similar molecular ground-state preparation protocols have been experimentally demonstrated in related works~\cite{liu_performing_2023,cao_emulating_2024,baldazzi_fourqubit_2025}.
Additional levels in qudits can also be exploited to reduce resource requirements in simulations of non-Hermitian dynamics arising in open quantum systems~\cite{kazmina_demonstration_2024,li_programmable_2025}.\\

Across these application areas, proof-of-concept demonstrations have consistently shown that qudits enable substantial reduction in entangling-gate counts.
Combined with the greater flexibility in designing application-specific qudit entangling operations, these results illustrate how qudits can ``squeeze more computational juice'' from near-term quantum hardware.

\begin{table}
    \centering
    {\rowcolors{1}{gray!12}{white}
    \begin{tabular}[c]{ |L{2cm}|L{2cm}|L{3cm}|L{3cm}|L{2cm}|L{3cm}|L{2cm}| }
        \hline
        \textbf{Simulation} & \multicolumn{2}{l|}{\textbf{High-energy physics}} & \textbf{Information scrambling} & \textbf{Open quantum systems} & \textbf{Condensed matter multi-level systems} & \textbf{Electronic structure} \\
        \cline{2-3}
        & \textbf{Neutrino} & \textbf{Lattice gauge theory} & & & & \\
        \hline
        Target system & \includegraphics[width=1\linewidth]{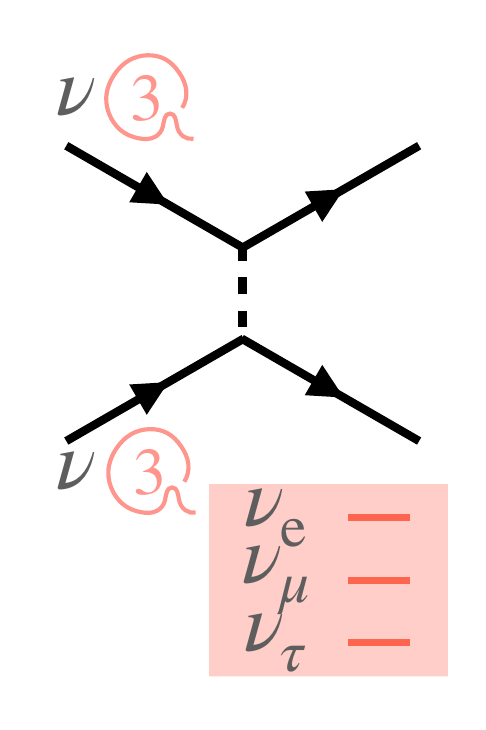} & \includegraphics[width=1\linewidth]{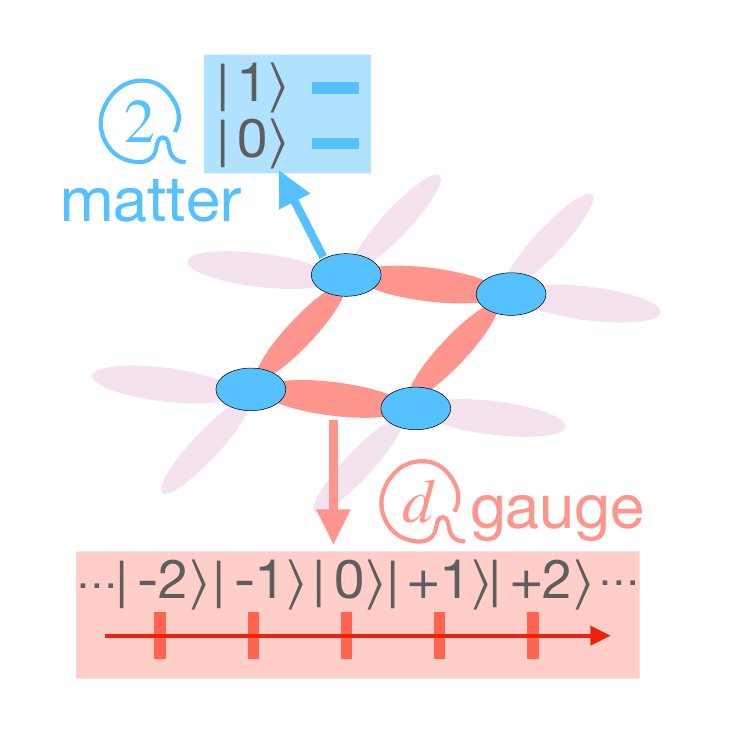} & \includegraphics[width=1\linewidth]{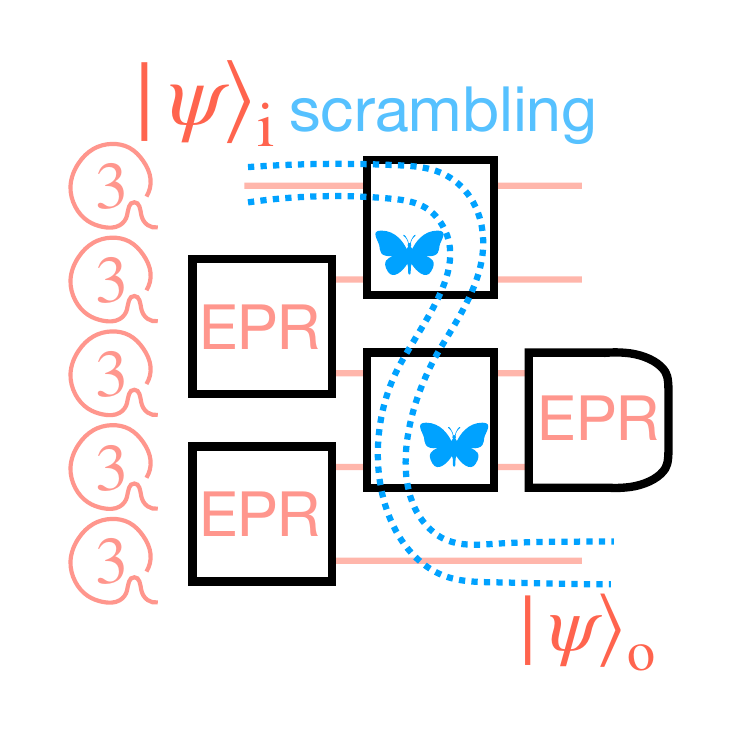} & \includegraphics[width=1\linewidth]{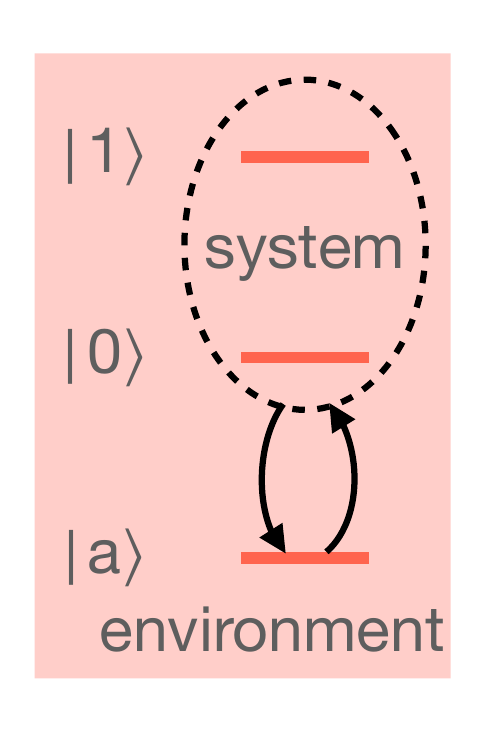} & \includegraphics[width=1\linewidth]{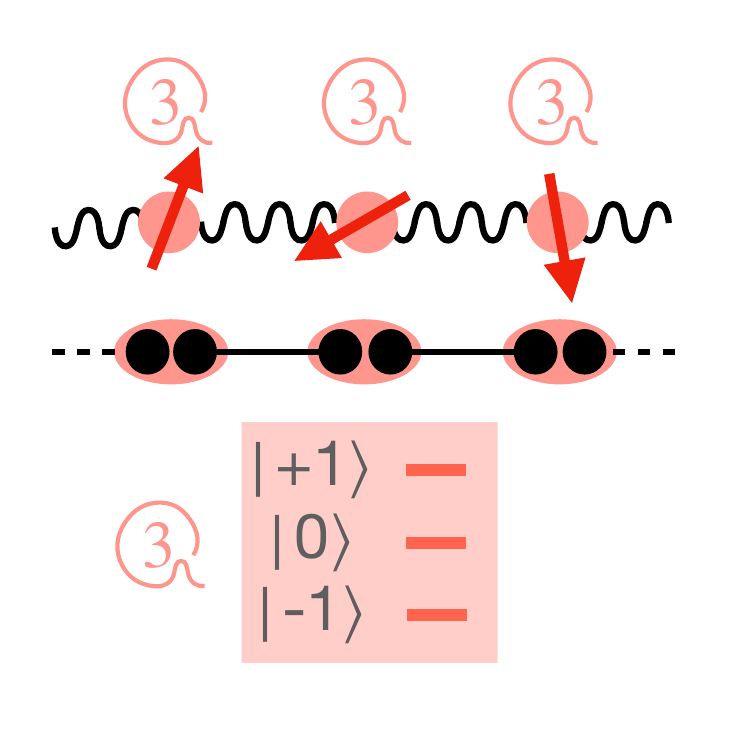} & \includegraphics[width=1\linewidth]{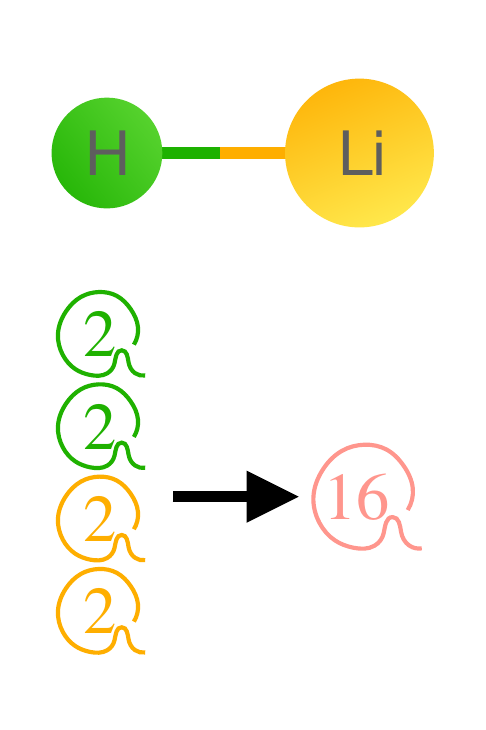}\\
        \hline
        Platform & SC circuits \cite{nguyen_simulating_2023,spagnoli_collective_2025} & Trapped ions \cite{meth_simulating_2025,john_nonabelian_2026}, SC circuits \cite{asaduzzaman_hybrid_2026} & SC circuits~\cite{blok_quantum_2021} & Trapped ions \cite{kazmina_demonstration_2024,li_programmable_2025}, SC circuits~\cite{kazmina_demonstration_2024} & Trapped ions \cite{senko_realization_2015,edmunds_symmetryprotected_2025,camacho_observing_2024}, SC circuits \cite{kumaran_transmon_2025,ticea_observation_2025,goss_qutrit_2026}, Rydberg atoms~\cite{qiao_realization_2025} & SC circuits \cite{liu_performing_2023,cao_emulating_2024}, Photons~\cite{kim_quditbased_2024,baldazzi_fourqubit_2025} \\
        \hline
        $n$ and~$d$ & $n=2,d=3$ \cite{spagnoli_collective_2025} & $n=2,d=3$~\cite{asaduzzaman_hybrid_2026}, $n=3,d=5$~\cite{meth_simulating_2025}, $n=3,d=8$~\cite{john_nonabelian_2026} & $n=5,d=3$~\cite{blok_quantum_2021} & $n=1,d=3$ \cite{kazmina_demonstration_2024}, $n=1,d=4$ \cite{li_programmable_2025} & $n=4,d=3$ \cite{senko_realization_2015,camacho_observing_2024,kumaran_transmon_2025}, $n=5,d=3$~\cite{edmunds_symmetryprotected_2025}, 
        $n=15,d=3$~\cite{goss_qutrit_2026},
        $n=25,d=3$~\cite{qiao_realization_2025}, $n=36,d=3$~\cite{ticea_observation_2025} & $n=1,d=4$ \cite{liu_performing_2023,cao_emulating_2024}, $n=1,d=16$ \cite{kim_quditbased_2024}, $n=2,d=4$ \cite{baldazzi_fourqubit_2025} \\
        \hline
        Entangling gates & Yes & Yes & Yes & No & Yes & No (Entangled photon pairs~\cite{baldazzi_fourqubit_2025}) \\
        \hline
    \end{tabular}
    }
    \caption{{\bf Quantum simulation using qudits.} The table presents a non-exhaustive overview of experimentally demonstrated quantum simulation examples across different research areas; additional examples are discussed in Sec.~\ref{subsec:simulation}. The mapping between the physical degrees of freedom in the target system and the qudits in quantum simulators is shown in shaded boxes where applicable.}
    \label{tab:simulation}
\end{table}

\subsection{Quantum computation}\label{subsec:computation}

It is much more challenging to build a universal quantum computer capable of executing arbitrary algorithm than to simulate the physics of a particular system. Unlike quantum simulation, universal quantum computation does not generally benefit from a natural problem-to-hardware mapping; furthermore, the asymptotic computational complexity of an algorithm is independent of the choice of base used to represent the problem. Nevertheless, qudit encodings often enable more resource-efficient implementations, reducing entangling-gate overhead and circuit depths. Enabled by recent advances in qudit control, several well-known quantum algorithms have been experimentally demonstrated on qudit platforms (see Table~\ref{tab:algorithms}).

Representative examples include Grover’s algorithm for unstructured database search~\cite{grover_fast_1996} and the quantum phase estimation algorithm~\cite{kitaev_quantum_1995}, a key subroutine of Shor’s factorisation algorithm~\cite{shor_polynomialtime_1997}. Shor’s algorithm offers an sub-exponential speedup for factoring large integers into prime factors, implying the ability to break widely used public-key encryption schemes. However, current quantum hardware remains far from the scale required for such applications, and experimental demonstrations are limited to very small problem sizes due to noise and gate infidelities. Qudit-based encodings offer a potential pathway toward larger problem instances by reducing the number of logical operations required. As a proof of concept for reduced register size and gate depth, Shor’s algorithm has been demonstrated on a two-ququart photonic quantum device~\cite{chi_programmable_2022}, factoring the number $15$, still a very small integer.
Scaling such approaches remains challenging.

Grover’s search algorithm provides a quadratic speedup over classical search methods and has been demonstrated across multiple qudit platforms, including trapped ions~\cite{shi_efficient_2026}, superconducting transmons~\cite{liu_performing_2023, roy_twoqutrit_2023} and nuclear-spin qudits~\cite{godfrin_operating_2017}. The largest database size that has been realised for the Grover algorithm is $N=9$ encoded into two transmon qutrits with entangling capabilities~\cite{roy_twoqutrit_2023}. For a single qudit, the largest database size demonstrated is $N=8$ on a trapped-ion device~\cite{shi_efficient_2026}.

Overall, these demonstrations primarily serve as proof-of-principle benchmarks, illustrating that current qudit control capabilities enable small-scale implementations of canonical quantum algorithms. In Table~\ref{tab:algorithms}, we summarise the system sizes and entangling capabilities of the qudit-based algorithmic demonstrations reported so far.
Various quantum algorithms are chosen to demonstrate implementation of qudit processing beyond characterising components.

Finally, we note that access to qudit operations has the potential to benefit the execution of algorithms written in terms of qubit gates. In particular, many algorithms, such as Grover's or Shor's, that employ subroutines based on classical Boolean logic or arithmetic employ $N$-body Toffoli gates. Since quantum hardware often does not have access to native $N$-body entangling operations, it is standard for $N$-body qubit gates to be decomposed into the native two-qubit gates present on the hardware, often incurring a large overhead in gate operations. Notably, accessing qudit-based entangling operations and employing auxiliary levels within the qudit state space as built-in ancillae can significantly reduce the number of two-body entangling operations required to generate an $N$-body qubit gate, with asymptotic advantages demonstrated in the large-$N$ limit~\cite{ralph_efficient_2007,lanyon_simplifying_2009,gokhale_asymptotic_2019,nikolaeva_decomposing_2022,nikolaeva_generalized_2023}. On the measurement side, the use of auxiliary levels can also aid in characterising and suppressing qubit readout errors~\cite{elder_highfidelity_2020,chen_transmon_2023,chen_enhancing_2025}, and implementing generalised qubit measurements described by positive operator-valued measures~\cite{fischer_ancillafree_2022,stricker_experimental_2022,suprano_experimental_2024,zia_quantum_2025}.

\begin{table}
    \centering
    {\rowcolors{1}{gray!12}{white}

\begin{tabular}[c]{ |L{3cm}|L{3cm}|L{3cm}|L{3cm}|}              
\hline
Algorithm & Platform & $n$ and~$d$ & Entangling gates \\
\hline
Deutsch--Jozsa algorithm~\cite{deutsch_rapid_1992,cleve_quantum_1998} & Photons~\cite{chi_programmable_2022}; SC circuits~\cite{roy_twoqutrit_2023} & \cite{chi_programmable_2022}:  $n=2, d=4$;  \cite{roy_twoqutrit_2023}: $n=2, d=3$ & Yes \\ 
\hline
Bernstein--Vazirani algorithm~\cite{bernstein_quantum_1993} & Photons~\cite{chi_programmable_2022}; Trapped ions~\cite{low_quantum_2025} & \cite{chi_programmable_2022}: $n=2, d=4$;  \cite{low_quantum_2025}: $n=1, d=8$ & \makecell[tl]{\cite{chi_programmable_2022}: Yes\\ \cite{low_quantum_2025}: No } \\
\hline
Phase estimation algorithm~\cite{kitaev_quantum_1995} & Photons \cite{chi_programmable_2022,lu_quantum_2020} & \cite{chi_programmable_2022}: $n=2, d=4$ & Yes \\
\hline
Shor's algorithm~\cite{shor_polynomialtime_1997} & Photons~\cite{chi_programmable_2022} & \cite{chi_programmable_2022}: $n=2, d=4$ used to factorise 15 & Yes \\
\hline
Grover's algorithm~\cite{grover_fast_1996} & Solid-state nuclear spin~\cite{godfrin_operating_2017}; SC circuits~\cite{roy_twoqutrit_2023,liu_performing_2023};
Trapped ions~\cite{shi_efficient_2026} & \cite{godfrin_operating_2017}: $n=1, d = 4$; \cite{roy_twoqutrit_2023}: $n=2, d=3$; \cite{shi_efficient_2026}: $n=1, d = 8$ &
\makecell[tl]{\cite{godfrin_operating_2017}: No \\ \cite{roy_twoqutrit_2023}: Yes \\ \cite{shi_efficient_2026}: No }\\
\hline
Permutation parity check algorithm~\cite{gedik_computational_2015} & NMR~\cite{gedik_computational_2015,dogra_determining_2014}; SC circuits~\cite{liu_performing_2023,yu_efficient_2025} & \cite{gedik_computational_2015,liu_performing_2023}: $n=1, d = 4$ &
No\\
\hline
\end{tabular}
}
\caption{{\bf Demonstrations of quantum algorithms on qudit platforms.} This non-exhaustive list highlights experimental implementation of representative quantum algorithms across different platforms, together with the corresponding Hilbert-space size used to realise it, characterised by $(d,n)$ (see Sec.~\ref{subsec:computation} for details).}
\label{tab:algorithms}

\end{table}

\subsection{Error suppression}
\label{sec:errorsuppression}
Qudits offer compelling advantages over qubits, but realising these benefits depends critically on device performance. Importantly, due to advantages such as reduced compilation overhead, qudit operations need not achieve comparable gate fidelities but must still pass a viable threshold~\cite{gokhale_asymptotic_2019, jiang_nonabelian_2025, fischer_universal_2023, heyfron_quantum_2019}.
Transcending from qubits to qudits complicates error mechanisms. Qudits rely on coherence between multiple energy levels, thus becoming simultaneously sensitive to more sources of both phase decoherence and energy relaxation even in a single physical platform, such as neutral atoms, trapped ions, or superconducting circuits. Moreover, due to higher degree of connectivity in coupled qudit systems, multi-body effects such as both classical and quantum crosstalk are more challenging to address as the qudit dimension increases~\cite{tripathi_qudit_2025}. Consequently,
managing and suppressing errors in qudit platforms is a central challenge for scalable computing.

Broadly speaking, error-suppression strategies for qudits can be grouped into three categories. The first category involves circuit-level suppression techniques, including dynamical decoupling~\cite{blok_quantum_2021,yuan_preserving_2022,zalivako_continuous_2023,iiyama_quditgeneralization_2024,tripathi_qudit_2025, zhou_robust_2024,read_dynamical_2025,read_platonic_2025}, robust pulse shaping~\cite{seifert_exploring_2023, seifert_timeefficient_2022, omanakuttan_qudit_2023, cho_direct_2024,basyildiz_speed_2025}, and specialised subroutines such as leakage removal units~\cite{mcewen_removing_2021}. These techniques improve performance at the control level, specifically addressing challenges such as decoherence and crosstalk, without fundamentally altering computational architecture. Notably, these techniques are applicable for improving qudit performance in both near-term applications and for longer-term fault-tolerant protocols.

The second category of error-suppression strategies focuses on error mitigation through post-processing. These techniques estimate noiseless expectation values from noisy circuit executions, typically by deliberately amplifying noise in a controlled manner and extrapolating back to the zero-noise limit~\cite{giurgica-tiron_digital_2020,ferracin_efficiently_2024}. Such sampling-based methods aim to improve performance of imperfect devices and are appropriate for near-term gate-based quantum simulation tasks~\cite{kim_evidence_2023}. Another type of error mitigation is randomised compiling, which involves performing many logically equivalent copies of a circuit with randomly selected local gates to significantly simplify a device's noise model~\cite{hashim_randomized_2021}. Qudit systems possessing a gate set similar to conventional qubit devices, i.e. a Clifford two-qudit gate and arbitrary single-qudit gates, can benefit from straightforward generalisations of qubit error mitigation techniques, often with no additional sampling overhead~\cite{goss_extending_2024}.

Finally, truly scalable and arbitrarily reliable qudit computation is expected to rely on logical encoding through quantum error correction~\cite{gottesman_introduction_2024}.
By embedding logical qudits into redundantly encoded states of many physical qudits, stabiliser-based codes provide a pathway toward fault-tolerant computation, often realising memories with higher code thresholds than their analogous qubit counterparts~\cite{watson_fast_2015,anwar_fast_2014,keppens_qudit_2025, spencer_qudit_2025}. In the context of fault-tolerant computing, universal qudit operation can be achieved with established techniques such as lattice surgery~\cite{cowtan_qudit_2022} and through highly efficient magic state distillation routines~\cite{sharma_magic_2024, prakash_magic_2020, campbell_magicstate_2012, campbell_enhanced_2014, sharma_nearthreshold_2024}.
A physical qudit Hilbert space can also be used to encode a logical qubit or qudit, which can itself be employed within a larger error-correcting code. For instance, using bosonic codes, the dephasing errors due to magnetic field fluctuations in Zeeman-encoded trapped-ion qudits can be converted into excitations of the ion motional mode and autonomously dissipated, leading to reduced error rates~\cite{debry_error_2026,li_beating_2025}. As another example, tailoring amplitude-damping events to monitored non-computational levels allows photon loss to be converted to erasure errors on logical qubits, which can be corrected more efficiently than Pauli errors~\cite{wu_erasure_2022,kubica_erasure_2023}. Such qudit-based erasure qubits have been experimentally demonstrated in trapped ions~\cite{shi_longlived_2025,quinn_highfidelity_2026}, both transmon~\cite{putterman_hardwareefficient_2025,liu_hardwareefficient_2026} and fluxonium~\cite{liu_converting_2026} superconducting circuits, and neutral atoms~\cite{ma_highfidelity_2023,scholl_erasure_2023}. In yet another example, multiple physical qubits are encoded in individual photons, such that multi-photon entangled states realise reconfigurable graph states that can be used for error-protected logical qubit operations~\cite{vigliar_errorprotected_2021}.
Whereas qubit error correction has seen increasingly mature experimental demonstrations, the development of comparable frameworks for qudits remains at an early stage. Nonetheless, a beyond-break-even quantum memory in $d=3,4$ has been demonstrated using a qudit Gottesman-Kitaev-Preskill (GKP) encoding in the bosonic modes of a superconducting cavity~\cite{brock_quantum_2025}.

\section{Looking forward}\label{sec:outlook}

Binary encodings are not always the natural choice in quantum computing. For many tasks---especially quantum simulations in high-energy and condensed-matter physics---qudits are more natural because the underlying degrees of freedom are themselves high-dimensional (see Table~\ref{tab:simulation}).
In this picture, $d$ becomes a tunable parameter chosen to match both the hardware and the computational task.
Future quantum computers may also benefit from hybrid registers that combine qubits with qudits with $d>2$~\cite{daboul_quantum_2003,litteken_qompress_2023,mato_compression_2023,mato_mqt_2024}, an approach already shown to be advantageous for quantum simulation~\cite{meth_simulating_2025,zhou_demonstration_2025}.
Moreover, qudits can access classically intractable regimes far more quickly than qubits: the Hilbert space of $70$ qubits is matched by only $24$ qudits with $d=8$. Importantly, they can also greatly reduce circuit complexity, with gate-depth savings of more than one order of magnitude demonstrated in certain quantum simulation tasks~\cite{meth_simulating_2025,john_nonabelian_2026}.
In addition, qudits offer a richer gate landscape than qubits, opening new opportunities for error-correcting codes and algorithm design tailored to higher dimensions.

The examples in Tables~\ref{tab:platforms}, \ref{tab:simulation} and~\ref{tab:algorithms} remain early NISQ-era demonstrations, but they already highlight how qudits can extract more computational power from quantum hardware. The ultimate goal is to scale these first advances into fault-tolerant quantum computers.
Companies such as IBM Quantum, Google, Quantinuum, Rigetti, Nord Quantique, Oxford Quantum Circuits, Pasqal, and Quantum Computing Inc. are already exploring this direction, and IBM's Qiskit~\cite{javadi-abhari_quantum_2024} and Google's Cirq~\cite{cirqdevelopers_cirq_2025} among other software toolkits support qudit operations.
Along this path, the realisation of high-quality qudit entangling gates is a key element.
Realising the full potential of qudits will require a broader toolbox for benchmarking and performance evaluation. Once quantum devices surpass the reach of classical verification, trusted qudit computation will depend on robust methods for validation and verification~\cite{hashim_practical_2025}.
Interconnects for quantum state transfer will also be important~\cite{shalaev_photonic_2025,thomas_highdimensional_2025}. They offer a scalable path by linking smaller quantum modules and, in the longer term, may also support a future quantum internet~\cite{nguyen_quantum_2024}.

It will be interesting to explore computing paradigms beyond the standard gate-based model. Examples include dissipative quantum computing~\cite{verstraete_quantum_2009,reiter_dissipative_2017,harrington_engineered_2022}, which relies on non-unitary dynamics, and measurement-based quantum computing~\cite{briegel_measurementbased_2009}, which proceeds through sequences of measurements. Both approaches have unique potential as stand-alone computing architecture or by incorporating elements in a gate-based architecture~\cite{chan_measurementbased_2024}, yet are currently underexplored for $d>2$.

Qudits impact not only quantum computing, but also quantum technologies such as quantum communication~\cite{cozzolino_highdimensional_2019} and quantum sensing~\cite{shlyakhov_quantum_2018,kristen_amplitude_2020,hesselmeier_quditbased_2024,zhou_robust_2024}. Higher‑dimensional entanglement improves quantum channel capacity, communication security and noise resilience; moreover, qudit sensors enhance measurement precision and allow more freedom in sensing protocol design.

In summary, qudit quantum computing is just beginning. Nature offers us qudits in many physical systems---the task ahead is to learn how to use them to full advantage.

\section*{Acknowledgements}
We thank Martin Ringbauer for critical reading of the manuscript.
We acknowledge support from the Natural Sciences and Engineering Research Council of Canada (NSERC), the Canada First Research Excellence Fund (CFREF, Transformative Quantum Technologies), New Frontiers in Research Fund (NFRF), Ontario Early Researcher Award, and the Canadian Institute for Advanced Research (CIFAR).

\clearpage
\bibliographystyle{myapsrev2}
\bibliography{biblio}

\end{document}